%% file: AoI_RAWNET.tex
\documentclass[journal]{IEEEtran}

\makeatletter
\def\ps@headings{%
\def\@oddhead{\mbox{}\scriptsize\rightmark \hfil \thepage}%
\def\@evenhead{\scriptsize\thepage \hfil \leftmark\mbox{}}%
\def\@oddfoot{}%
\def\@evenfoot{}}
\makeatother
\usepackage[pdftex]{graphicx}
\usepackage{amsmath}
\usepackage{amssymb}
\usepackage{dblfloatfix}
\usepackage{array}
\usepackage[svgnames]{xcolor} 
\usepackage{mdwmath}
\usepackage{subfigure}
 \usepackage{bm}
\usepackage{lettrine}
\usepackage{tipa}
 \usepackage{hyperref}
 \usepackage{framed}
 \usepackage{booktabs}
 \usepackage{caption}
 \usepackage{url}

\usepackage{algorithm}
\usepackage{algorithmic}
\usepackage{epsfig}
\usepackage{overpic}
\usepackage{epstopdf}
\usepackage{amsfonts,setspace,amsthm,braket}

\graphicspath{{../pdf/}{../jpeg/}}
\DeclareGraphicsExtensions{.pdf,.jpeg,.png}

\usepackage{dsfont}
\newtheorem{theorem}{Theorem}

\newtheorem{proposition}[theorem]{Proposition}

\allowdisplaybreaks

\newcommand\EatDot[1]{}
\DeclareMathOperator*\lowlim{\underline{lim}}
\DeclareMathOperator*\uplim{\overline{lim}}
\newcommand{\edit}[1]{{ #1}}

\floatname{algorithm}{Algorithm}
\IEEEoverridecommandlockouts

\begin{document}
\title{On Minimizing the Maximum Age-of-Information For Wireless Erasure Channels}

\author{
\IEEEauthorblockN{Arunabh Srivastava, Abhishek Sinha, Krishna Jagannathan\\}
\IEEEauthorblockA{Dept. of Electrical Engineering, IIT Madras\\}
Email: \{ee16b132, abhishek.sinha, krishnaj\}@ee.iitm.ac.in
}
\maketitle

\input{abstract}
\input{introduction}

\input{model}

\input{max_AoI}
\input{throughput}

\input{simulations}

\input{conclusion}
\input{Acknowledgement}
\bibliographystyle{IEEEtran}
\bibliography{MIT_broadcast_bibliography}
\input{appendix}

\end{document}

%% file: abstract.tex
\begin{abstract}
  \emph{Age-of-Information} (AoI) is a recently proposed metric for quantifying the freshness of information from the UE's perspective in a communication network. Recently, Kadota et al. \cite{AoI_ToN} have proposed an index-type approximately optimal scheduling policy for minimizing the \emph{average}-AoI metric for a downlink transmission problem. For delay-sensitive applications, including real-time control of a cyber-physical system, or scheduling URLLC traffic in 5G, it is essential to have a more stringent \emph{uniform} control on AoI across all users. In this paper, we derive an exactly optimal scheduling policy for this problem in a downlink cellular system with erasure channels. Our proof of optimality involves an explicit solution to the associated average-cost Bellman Equation, which might be of independent theoretical interest. We also establish that the resulting Age-process is positive recurrent under the optimal policy, and has an exponentially light tail, with the optimal large-deviation exponent. Finally, motivated by typical applications in small-cell residential networks, we consider the problem of minimizing the peak-AoI with throughput constraints to specific UEs, and derive a heuristic policy for this problem. Extensive numerical simulations have been carried out to compare the efficacy of the proposed policies with other well-known scheduling policies, such as Randomized scheduling and Proportional Fair. 
\end{abstract}

\begin{IEEEkeywords}
Age-of-Information, Scheduling, Bellman Equation, Stability. 	
\end{IEEEkeywords}

%% file: introduction.tex
\section{Introduction}
\lettrine[]{\textbf{K}}{ }eeping information fresh is an essential requirement for a variety of control and communication tasks. Stale feedback information in networked control systems may reduce the gain or phase margin, which may, in turn, push the system towards the verge of instability \cite{Johari},\cite{instability}. Real-time status updates are necessary for a plethora of communication tasks including effective traffic monitoring \cite{ferrier1994real}, online gaming \cite{neumann2007challenges}, intrusion detection \cite{Dousse}, environment sensing using IoT devices \cite{schulz2017latency} etc. While designing routing and scheduling policies for maximizing the throughput region is well-understood \cite{tassiulas}, \cite{sinha_umw}, designing optimal policies for maximizing the information freshness is currently an active area of research \cite{joo2018wireless}.  

With the advent of the 5G technology, it is becoming increasingly common for the Base Stations (BS) to serve the following two different types of UEs at the same time - \textbf{Type-I:} Delay-constrained UEs (\emph{e.g.}, UEs with URLLC type of traffic \cite{popovski2017ultra}, such as control information updates for autonomous vehicles), and \textbf{Type-II:} Throughput-constrained UEs (\emph{e.g.,} UEs with eMBB type of traffic, such as HD multimedia streaming). Moreover, in Network-Control applications where the delay-constrained UEs often perform a global task jointly (\emph{e.g.}, by sensing different parts of a sizeable cyber-physical system), it is critical to \emph{uniformly} maximize the information freshness across all Type-I UEs to avoid information bottlenecks. In this paper, we study the problem of optimal joint scheduling of Type-I and Type-II UEs over wireless erasure channels. 

We characterize the freshness of information at a UE by a metric called the \emph{Age of Information} (AoI) \cite{kaul2012real}, \cite{kosta2017age}. In our context, the AoI for a UE at a time $t$ indicates the time elapsed since the UE received a new packet from the BS prior to time $t$. The larger the value of AoI for a UE at a time, the more outdated the UE is at that time. In this short paper, we consider two related problems on minimizing the AoI - \textbf{Problem (1):} In the presence of only Type-I UEs, our goal is to design a scheduling policy which minimizes the long-term \emph{peak}-AoI \emph{uniformly} across all UEs, and \textbf{Problem (2):} When the Type-II UEs are to be scheduled simultaneously along with the Type-I UEs by the same BS, we consider the problem of minimizing the long-term \emph{peak}-AoI, subject to throughput constraint to the Type-II UEs. 

\paragraph*{Related Work} 
In a recent paper \cite{bedewy2018age} Bedewy et al., consider the problem of optimal scheduling of status updates over an error-free delay channel. They showed that, in that setting, the greedy \emph{Max-Age First} scheduling policy is an optimal policy for both peak-age and total age metrics. In the paper \cite{AoI_Sched}, He et al. consider the problem of link scheduling to transmit a fixed number of packets over a common interference-constrained channel such that the overall age is minimized. They proved the problem to be NP-hard and proposed an Integer Linear Program and a fast heuristic. The authors continued studying the previous problem in \cite{AoI_peak} for minimizing the peak-age, and obtained similar results. See the monograph \cite{kosta2017age} for a detailed survey of the recent literature on Age of Information.

Closer to our work is the paper \cite{AoI_ToN}, which studies a similar problem with single-hop wireless erasure channel. However, contrary to this paper, the objective of \cite{AoI_ToN} is to design a policy to minimize the long-term \emph{average} AoI. Using Lyapunov-drift based methodology, the paper \cite{AoI_ToN} designs an \emph{approximately optimal} policy for this problem. Designing an  optimal policy in this setting still remains elusive. As we argued before, with distributed sensing applications, where \emph{all} sensors need to stay updated uniformly, a more suitable objective is to minimize the long-term \emph{peak}-AoI across all users. In this paper, we design an \emph{exactly optimal} policy for the problem (1) using MDP techniques. \edit{We also show that the proposed policy achieves the optimal large deviation exponent among all scheduling policies.} Moreover, inspired by the analysis for the problem (1), we propose a heuristic policy for the problem (2), where we incorporate an additional throughput constraint for the eMBB UEs. Operating performances of these proposed policies have been compared extensively with other well-known scheduling policies through numerical simulations. \\
The rest of the paper is organized as follows. Section \ref{system_model} outlines the system model. In Section \ref{max_AoI}, we consider the problem of minimizing the long-term peak-AoI across all UEs when only Type-I UEs are present. In Section \ref{throughput}, we consider the problem of optimal joint scheduling in the presence of both Type-I and Type-II UEs. Section \ref{simulation} presents numerical simulation results comparing the proposed policies with other well-known scheduling policies, such as Proportional Fair and Randomized Policies. Finally, we conclude the paper in Section \ref{conclusion}.

%% file: model.tex
\section{System Model} \label{system_model}
\begin{figure}
\vspace{-60pt}
\hspace{-40pt}
\centering
\begin{overpic}[width=0.38\textwidth]{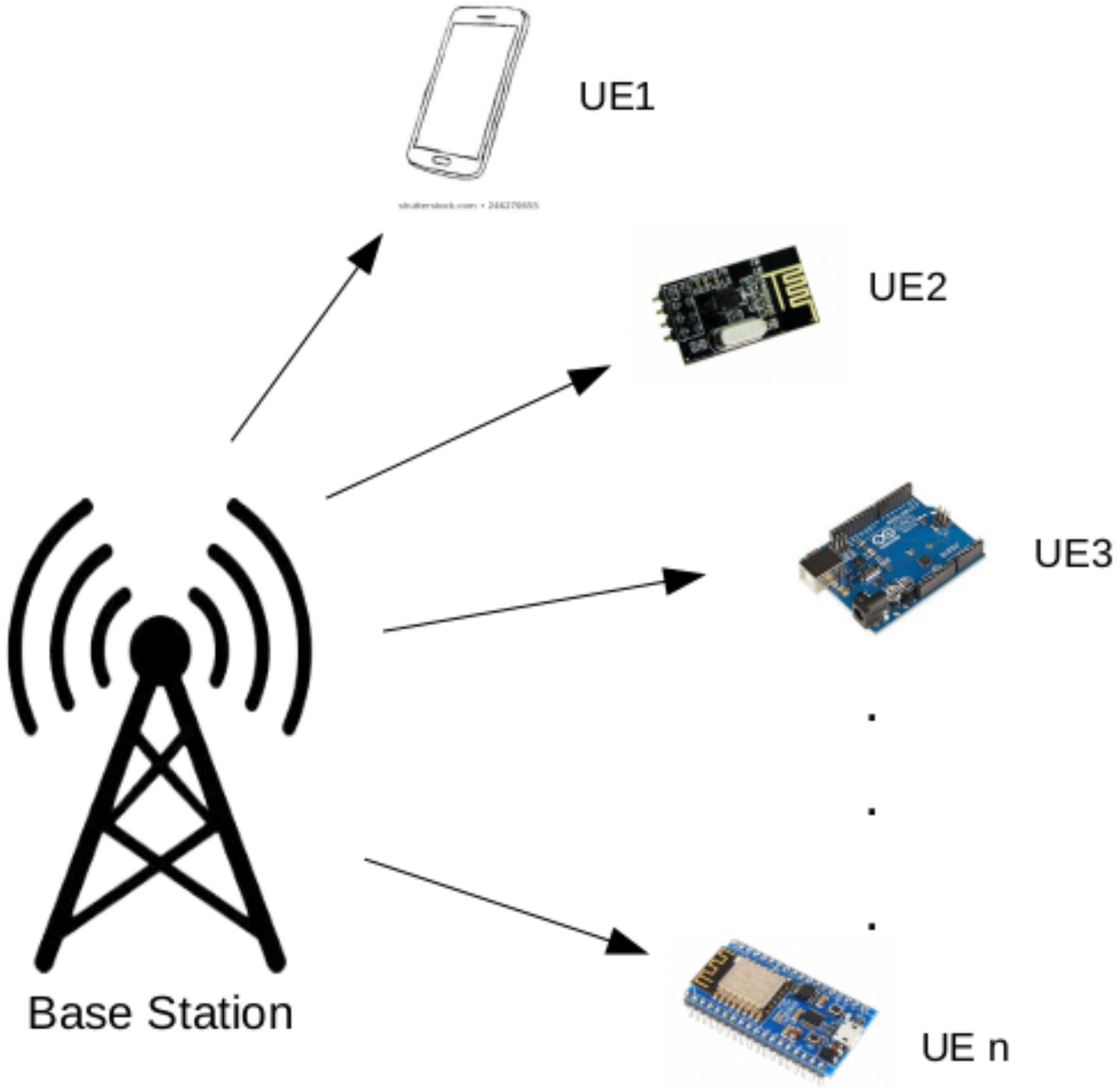}
\put(45,76){\footnotesize {Type-II UE}}
\put(45,72){\footnotesize{(eMBB Device)}}
\put(21, 60){$p_1$}
\put(31, 57){$p_2$}
\put(35, 48){$p_3$}
\put(36, 33){$p_n$}
\put(68,45){\Bigg \}}
\put (72, 52){\footnotesize{Type-I UEs}}
\put (72, 47){\footnotesize{(URLLLC}}
\put (73, 44){\footnotesize{Devices)}}
\end{overpic}
\vspace{-40pt}
\caption{\small {A Base Station serving packets to an eMBB device (UE\textsubscript{1}) and multiple URLLC devices (UE\textsubscript{2}-UE\textsubscript{n}) over a wireless erasure channel.}}
\label{fig:model}
\end{figure}

We consider the downlink UE scheduling problem where a Base Station (BS) serves $N$ wireless users, each with \emph{full-buffer} traffic, meaning,  each user is infinitely backlogged. The channel from the BS to the $i$\textsuperscript{th} UE is modelled by a binary erasure channel with erasure probability $1-p_i$, where $p_i>0, \forall i$. Time is slotted, and the BS can transmit to only one user per slot. If the BS transmits a packet to the $i$\textsuperscript{th} UE at slot $t$, the packet is either successfully decoded by the UE with probability $p_i$, or, the packet is permanently lost with probability $1-p_i$, independently of everything else. Refer to Figure \ref{fig:model} for a schematic diagram of the model. The objective is to design suitable downlink UE scheduling policy optimizing  a given metric. In this paper, we consider two related problems - (1) designing peak-AoI-optimal scheduler without any throughput constraints (Section \ref{max_AoI}) and (2) Designing peak-AoI-optimal scheduler with throughput constraint for a UE (Section \ref{throughput}). 

%% file: max_AoI.tex
\section{Minimizing the Peak-Age-of-Information}\label{max_AoI}
In this section, we consider the problem of minimizing the long-term \emph{peak}-AoI metric, which denotes the \emph{maximum} Age of Information among all receivers associated with a base-station in a cellular network. With hard deadline constraints for each user in the case of URLLC traffic in 5G, minimizing the \emph{peak}-AoI metric is more practically meaningful than the minimizing the \emph{average} AoI \cite{AoI_ToN}, \cite{AoI_infocom}. \\
At a given slot $t$, define $h_{\max}(t)\equiv \max_{i=1}^{N} h_i(t)$ to be the peak \emph{instantaneous} Age of Information among all $N$ users. Our objective is to design a scheduling policy $\pi^* \in \Pi$, which minimizes the time-averaged expected peak-AoI. More formally, we consider the following stochastic control problem $\mathcal{P}_{\textrm{sched}}$:
\begin{equation} \label{problem}
\lambda^*= \inf_{\pi \in \Pi} \limsup_{T\to \infty} \frac{1}{T}\sum_{t=1}^{T} \mathbb{E}(\max_i h_i(t)),
\end{equation}
subject to the constraint that at most one user may be scheduled at any slot. 
Define a greedy scheduling policy \textsf{MA} (Max-Age) which, at any given slot $t$, schedules the user having the \emph{highest} instantaneous age. More formally, at a slot $t$, a user $i$ is scheduled which maximizes the metric $\max_{i=1}^{N} h_i(t)$ (ties are broken arbitrarily). We establish the following theorem for \textsf{MA}:

\begin{framed}
\begin{theorem}[Optimal Policy] \label{opt}
	The greedy policy \textsf{MA} is an optimal policy for the problem \eqref{problem}. Moreover, the optimal long-term peak-AoI is given by $\lambda^*=\sum_{i=1}^{N} \frac{1}{p_i}$.
\end{theorem}
\end{framed}
Theorem \ref{opt} states that the greedy policy \textsf{MA} is optimal for the peak-AoI metric. Interestingly, the optimal policy is independent of the channel statistics (the probability vector $\bm{p}$). This should be contrasted with the approximately optimal policy \textsf{MW} for minimizing the \emph{average}-AoI metric proposed in \cite{AoI_ToN}. \\
We prove this theorem by proposing a closed-form candidate solution of the Bellman's equation of the associated average-cost MDP and then verifying that the candidate solution indeed satisfies the Bellman's equation.
 \begin{IEEEproof}
 	The stochastic control problem under investigation is an instance of a countable-state average-cost MDP with a finite action space. The state $\bm{h}(t)$ of the system at a slot $t$ given by the current AoI vector of all users, i.e., $\bm{h}(t)\equiv (h_1(t),h_2(t), \ldots, h_N(t))$. The per-stage cost at time $t$ is $\max_{i=1}^{N}h_i(t)$, which is unbounded, in general. Finally, the finite action space $\mathcal{A}=\{1,2,\ldots, N\}$ corresponds to the index of the user scheduled at a given slot. \\
 	Let the optimal cost for the problem $\mathcal{P}_{\textrm{sched}}$ be denoted by $\lambda^*$ and the differential cost-to-go from the state $\bm{h}$ be denoted by $V(\bm{h})$. Then, following the standard theory of average cost countable state MDP (Proposition 4.6.1 of \cite{bertsekas2005dynamic}), we consider the following Bellman Eqn.  
 	\begin{eqnarray} \label{bellman}
 		&&\lambda^*+ V(\bm{h})\\
 		&=& \min_i\{ p_i V(\underbrace{1}_{i\textsuperscript{th} \textrm{ coordinate}}, \bm{h}_{-i}\bm{+1})+ (1-p_i) V(\bm{h+1})\} \nonumber \\
 		&+& \max_i h_i \nonumber, 
 	\end{eqnarray}
 	where the vector $\bm{h}_{-i}$ denotes the $N-1$ dimensional vector of all coordinates excepting the $i$\textsuperscript{th} coordinate and $\bm{1}$ is a all-one vector.
 	\paragraph*{Explanation} The Bellman Equation \eqref{bellman} may be explained as follows. Suppose that the current AoI state is given by $\bm{h}$. If the scheduler schedules a transmission to the $i$\textsuperscript{th} user, the transmission is successful with probability $p_i$ and is unsuccessful with probability $1-p_i$. If the transmission is successful, the AoI of all users, excepting the $i$\textsuperscript{th} user, is incremented by $1$, and the AoI of the $i$\textsuperscript{th} UE is reduced to $1$. This explains the first term. On the other hand, if the transmission to the $i$\textsuperscript{th} UE is unsuccessful, the AoI of all users are incremented by $1$. This explains the second term within the bracket. Finally, the term $\max_i h_i$ denotes the current stage cost.
 	   
\paragraph*{Solution to the Bellman Equation \eqref{bellman}} We verify that the following constitutes a solution to the Bellman Equation \eqref{bellman}:
\begin{eqnarray} \label{bellman_soln}
	V(\bm{h})= \sum_j \frac{h_j}{p_j}, \hspace{10pt}
	\lambda^*=\sum_j \frac{1}{p_j}.
\end{eqnarray} 	
 To verify the solution, we start with the $\mathsf{RHS}$ of \eqref{bellman}. Upon substitution from Eqn. \eqref{bellman_soln}, the expression corresponding to the $i$\textsuperscript{th} user inside the $\min$ operator of Eqn. \eqref{bellman} is simplified to: 
 \begin{eqnarray}
 	&&p_i V(1, \bm{h}_{-i}\bm{+1})+ (1-p_i) V(\bm{h+1}) \nonumber \\
 	&=& p_i \sum_{j\neq i} \frac{h_j+1}{p_j}+ 1 + (1-p_i) \sum_j \frac{h_j+1}{p_j}\nonumber \\
 	&=& p_i \sum_{j} \frac{h_j+1}{p_j} - p_i\frac{h_i+1}{p_i}+1 + (1-p_i) \sum_j \frac{h_j+1}{p_j}\nonumber \\
 	&=& \sum_j \frac{h_j}{p_j} - h_i + \sum_j \frac{1}{p_j}. \label{calc}
  \end{eqnarray}	
  Hence,
  \begin{eqnarray*}
  &&\mathsf{RHS}\\
  	&=&\min_i\{ p_i V(1, \bm{h}_{-i}\bm{+1})+ (1-p_i) V(\bm{h+1})\} + \max_i h_i \\
  	&=& \sum_j \frac{1}{p_j} + \sum_{j}\frac{h_j}{p_j} - \max_i h_i + \max_i h_i\\
  	&=& \lambda^* + V(\bm{h})\\
  	&=& \mathsf{LHS}.
  \end{eqnarray*}
 	
The optimality result now follows from \cite{bertsekas2005dynamic}.
\end{IEEEproof} 
The following interesting features of the optimal scheduling policy \textsf{MA} should be noted: 
\begin{itemize}
	\item Unlike the approximately optimal policy for the average-AoI metric proposed in \cite{AoI_ToN}, the optimal policy for the peak-AoI metric is \emph{completely agnostic} of the channel statistics parameter $\bm{p}$. Hence, the policy \textsf{MA} is simple to implement in practice as it requires no complex channel estimation procedures. 
	\item The proof of optimality of the \textsf{MA} policy gives an explicit expression for the associated cost-to-go function $V(\cdot)$ and the optimal cost $\lambda^*$. This is one of the rare cases where the associated Bellman Equation of an MDP has an analytic solution.   
\end{itemize}

\subsection{Large Deviation Rate Optimality and Stability}
\edit{Although Theorem \ref{opt} establishes that the \textsf{MA} scheduling policy is optimal in terms of minimizing the long-term \emph{expected} peak-AoI, for mission-critical URLLC applications, we need to additionally ensure that the peak-AoI metric stays within a bounded limit with high probability.} The following Proposition \ref{exp_bd} shows that the peak-Age process has an \emph{exponentially light tail} under the action of the \textsf{MA} policy. This ensures high-probability delay guarantees to URLLC traffic having a strict latency requirement. 
\begin{framed} 
\begin{proposition} \label{exp_bd}
	Under the action of the \textsf{MA} policy, there exists a constant $c(N, \bm{p})>0$ such that, for any fixed time $t\geq  1$ and any $k\geq 2N$, 
	\begin{equation} \label{ld_eqn}
	\mathbb{P}^{\textsf{MA}}(\max_i h_i(t) \geq k)\leq  c(N, \bm{p})k^N (1-p_{\min})^k.
	\end{equation}  
\end{proposition}
\end{framed}
See Appendix \ref{exp_bd_proof} for the proof of Proposition \ref{exp_bd}.\\


\edit{
 Our objective in the rest of this subsection is to show that the large deviation bound \eqref{ld_eqn} is asymptotically optimal in the sense that no other scheduling policy $\pi$ has lower probability exceeding a given sufficiently large AoI-threshold $k$. Towards this end, in the next proposition, we establish a fundamental performance bound of the peak-AoI tail probability under the action of \emph{any} scheduling policy.
\begin{framed}
\begin{proposition} \label{ld_lb_prop}
	Under the action of any arbitrary scheduling policy $\pi$, at any slot $t \geq k$ and for all $k \geq 1$, we have 
	\begin{eqnarray} \label{prop3:eq}
\mathbb{P}^\pi(\max_i h_i(t) \geq k) \geq (1-p_{\min})^k.
	\end{eqnarray}
\end{proposition}
\end{framed}
\begin{proof}
	 Let $i^* = \arg \min_i p_i$. Now,
	\begin{eqnarray*} \label{ld_lb}
		&&\mathbb{P}^\pi(\max_i h_i(t) \geq k) \geq \mathbb{P}^\pi(h_{i^*}(t) \geq k) \stackrel{(a)}{\geq} (1-p_{\min})^k, 
\end{eqnarray*}
where the inequality (a) follows from the fact that consecutive $k$ erasures just prior to time $t$ for UE\textsubscript{$i^*$} (which takes place with probability $(1-p_{\min})^k$) ensures that the age of UE\textsubscript{$i^*$} at time $t$ is at least $k$. 
\end{proof}
Combining Propositions \ref{exp_bd} and \ref{ld_lb_prop}, we conclude that the \textsf{MA} policy achieves the optimal large-deviation exponent for the peak-AoI metric.
\begin{framed}
\begin{theorem} \label{ld_opt}
	The \textsf{MA} policy achieves the optimal large-deviation exponent for the max-age metric and the value of the optimal exponent is given by 
	\begin{equation*}
		-\lim_{k\to \infty}\lim_{t\to \infty}\frac{1}{k}\log \mathbb{P}^{\textsf{MA}}(\max_i h_i(t) \geq k) = - \log(1-p_{\min}).
	\end{equation*}
\end{theorem}
\end{framed}
\begin{proof}
	From Proposition \ref{exp_bd}, since the inequality in Eqn. \eqref{ld_eqn} holds for any time $t\geq 1$, for the \textsf{MA} policy, we can write 
	\begin{eqnarray*}
		\uplim_{t \to \infty} \frac{1}{k}\log  \mathbb{P}^{\textsf{MA}}(\max_i h_i(t) \geq k)&\leq &  \frac{\log c(N, \bm{p})}{k}+ N \frac{\log k}{k}  \\
		&& +\log (1-p_{\min}).
	\end{eqnarray*}
	Hence, taking limit as $k \to \infty$, we obtain 
	\begin{eqnarray} \label{th4:eq0}
		\uplim_{k \to \infty}\uplim_{t \to \infty} \frac{1}{k}\log  \mathbb{P}^{\textsf{MA}}(\max_i h_i(t) \geq k)&\leq & \log(1-p_{\min}).
	\end{eqnarray}
On the other hand, for any fixed $k$ and at any time slot $t \geq k$, Eqn. \eqref{prop3:eq} of Proposition \ref{ld_lb_prop} states that for \emph{any} scheduling policy $\pi$, we have
\begin{eqnarray} \label{th4:eq1}
 	\frac{1}{k}\log \mathbb{P}^{\pi}(\max_i h_i(t) \geq k) \geq  \log (1-p_{\min}).
\end{eqnarray}
Since the bound \eqref{th4:eq1} is valid for any $t \geq k$, for any fixed $k \geq 0$, we can let $t \to \infty$ to obtain 
\begin{equation} \label{th4:eq2}
	\lowlim_{t \to \infty}\frac{1}{k}\log \mathbb{P}^{\pi}(\max_i h_i(t) \geq k) \geq  \log (1-p_{\min}).
\end{equation}
Finally, since the bound \eqref{th4:eq2} is valid for any $k \geq 0$, we can now let $ k \to \infty$ to obtain 
\begin{equation} \label{th4:eq3}
	\lowlim_{k \to \infty}\lowlim_{t \to \infty}\frac{1}{k}\log \mathbb{P}^{\pi}(\max_i h_i(t) \geq k) \geq  \log (1-p_{\min}).
\end{equation}
The proof now follows from Eqns \eqref{th4:eq0} and \eqref{th4:eq3}.
\end{proof}
\paragraph*{Discussion} Interestingly enough, although the time-average optimal performance depends on the statistical parameters of \emph{all} UEs (Theorem \ref{opt}), the optimal large-deviation exponent depends only on the parameter of the \emph{worst} UE.

}
We conclude this Section with the following stability result of the Age-process $\{\bm{h}(t)\}_{t\geq 1}$ under the \textsf{MA} policy. 
\begin{framed}
\begin{theorem} \label{stability}
The Markov Chain $\{\bm{h}(t)\}_{t \geq 1}$ is Positive Recurrent under the action of the \textsf{MA} policy. 
\end{theorem}	
\end{framed}
See Appendix \ref{stability_proof} for the proof. 


%

%% file: throughput.tex
\section{Minimizing the Peak-Age-of-Information with Throughput Constraints} \label{throughput}
In this Section, we consider a generalization of the above system model, where, in addition to maintaining a small peak-AoI, there is also a Type-II UE  (denoted by UE\textsubscript{$1$}), which is interested in maximizing its throughput. This problem can be motivated by considering a residential subscriber who is running one high-throughput application (with eMBB-type traffic), such as, downloading an HD movie, while also using several smart home automation IoTs, which have URLLC-type traffic, and hence, require low-latency. In a small-cell residential network, all of these devices are served by a single BS typically located within the house \cite{weitzen2013large}. 
\paragraph*{Objective} Define a sequence of random variables $\{\bar{a}_1(t)\}_{t\geq 1}$ such that $\bar{a}_1(t)=1$ if the UE\textsubscript{$1$} \emph{did not} successfully receive a packet at the end of slot $t$ and $\bar{a}_1(t)=0$ otherwise. Let $\beta \geq 0$ be a non-negative tuning parameter. We are interested in finding a scheduling policy which solves the following problem $\mathcal{P}'_{\textrm{sched}}$:
\begin{equation}\label{tput}
	\lambda^{**}= \inf_{\pi} \limsup_{T\to \infty} \frac{1}{T}\sum_{t=1}^{T} \mathbb{E}\bigg(\max_i h_i(t) +\beta \bar{a}_1(t)\bigg).
\end{equation}
The single-stage cost described above may be understood as follows: the first term $\max_i h_i(t)$ denotes the usual maximum AoI across all UEs as in the previous Section. The second term imposes a penalty of $\beta$ if the UE\textsubscript{1} does not receive a packet at the current slot. By suitably controlling the value of $\beta$, a tradeoff between the peak-AoI and achievable throughput to UE\textsubscript{1} (serving the eMBB traffic) may be obtained \cite{altman1999constrained}. \\
Similar to the problem $\mathcal{P}_{\textrm{sched}}$ of Eqn. \eqref{problem}, the problem $\mathcal{P}'_{\textrm{sched}}$ is also an instance of an infinite-state average-cost MDP with an additional action-dependent additive per-stage cost term ($\beta \bar{a}_1(t)$). Arguing as before, and introducing an additional cost term $g_i$ arising due to the throughput constraint, the Bellman Equation for this problem may be written down as follows:

\begin{align} \label{BE2}
	&\lambda^{**}+V(\bm{h})= \\
	 &\min_i\{ p_i V(1, \bm{h}_{-i}\bm{+1}) + (1-p_i) V(\bm{h+1})+g_i \}+ \max_i h_i, \nonumber
\end{align}
where
\begin{eqnarray} \label{cq}
	g_i=\begin{cases}
		\beta, \textrm{if} \hspace{5pt}i\neq 1  \\
		\beta (1-p_1), \textrm{if} \hspace{5pt} i=1.
	\end{cases}
\end{eqnarray}
Here $g_i$ denotes the expected cost when the UE\textsubscript{$1$}, which receives eMBB traffic, does not successfully receive a packet at slot $t$. The above Bellman Equation \eqref{BE2} may be explained along similar line as the equation \eqref{bellman}. 

Inspired by its similarity to the problem $\mathcal{P}_{\textrm{sched}}$, we try the same differential cost-to-go function $V(\bm{h})=\sum_i \frac{h_i}{p_i}$ as before. Using Eqn. \eqref{calc}, the RHS of Eqn. \eqref{BE2} can be evaluated to be 
\begin{eqnarray*}
	V(\bm{h}) + \sum_{j} \frac{1}{p_j} + \min_{i}\big(-h_i+g_i\big) + \max_i h_i. 
\end{eqnarray*}
This yields the \textsf{MATP} (Max-Age with Throughput) scheduling policy, which minimizes the RHS of the Bellman Equation \eqref{BE2}:
\begin{framed}
	\textsf{MATP}: At any slot $t$, schedule the user $i$ having the highest value of $h_i(t)-g_i$, where $g_i$ is given in Eqn. \eqref{cq}.
\end{framed}
The \textsf{MATP} policy strikes a balance between minimizing the peak-AoI (through the first term $h_i(t)$) while also ensuring sufficient throughput to the eMBB user (through the second term $g_i$). As $\beta$ is increased, it gradually dominates the AoI term, which, in turn, facilitates scheduling the eMBB user. 

\paragraph*{Analysis of \textsf{MATP}} Note that, 
\begin{equation} \label{aprox_bd}
\min_i\big( -h_i+g_i\big) \geq \min_{i} (-h_i) + \min_{i} g_i= -\max_i h_i + \beta (1-p_1).
\end{equation}
Furthermore, since $g_i \leq \beta, \forall i$, we have 
\begin{equation}
	\min_i (-h_i+g_i) \leq \min_i (-h_i + \beta ) = -\max_i h_i + \beta. 
\end{equation}
Hence, by taking $\lambda^{**}\equiv \sum_{j} \frac{1}{p_j} + \beta$, we see that under the action of the \textsf{MATP} policy, the sup-norm of the difference between the RHS and LHS of the Bellman Equation \eqref{BE2} is bounded by the constant $\beta p_1$. In other words, upon denoting the RHS of the Bellman operator of \eqref{BE2} by $T(\cdot)$ (see \cite{bertsekas2005dynamic} for this operator notation), we have 

\begin{equation}
	||V-TV||_\infty \leq \beta p_1.
\end{equation}
%
Hence, we conclude that the policy \textsf{MATP} approximately solves the Bellman Equation \eqref{BE2}. The efficacy of this policy is studied extensively in the Simulation Section \ref{simulation}.    

It should be noted that unlike the \textsf{MA} policy, the \textsf{MATP} policy takes into account the channel statistics (the value of $p_1$) for the eMBB user. It is oblivious to the channel statistics of other URLLC UEs, however. 

Following a similar line of argument as in the proof of Theorem \ref{stability}, we can establish the result below.

\begin{framed}
\begin{theorem} \label{stability_TP}
The Markov Chain $\{\bm{h}(t)\}_{t \geq 1}$ is Positive Recurrent under the action of the \textsf{MATP} policy. 
\end{theorem}	
\end{framed}

%% file: simulations.tex
\section{Numerical Simulation} \label{simulation}
\begin{figure}
\centering
\begin{overpic}[width=0.4\textwidth]{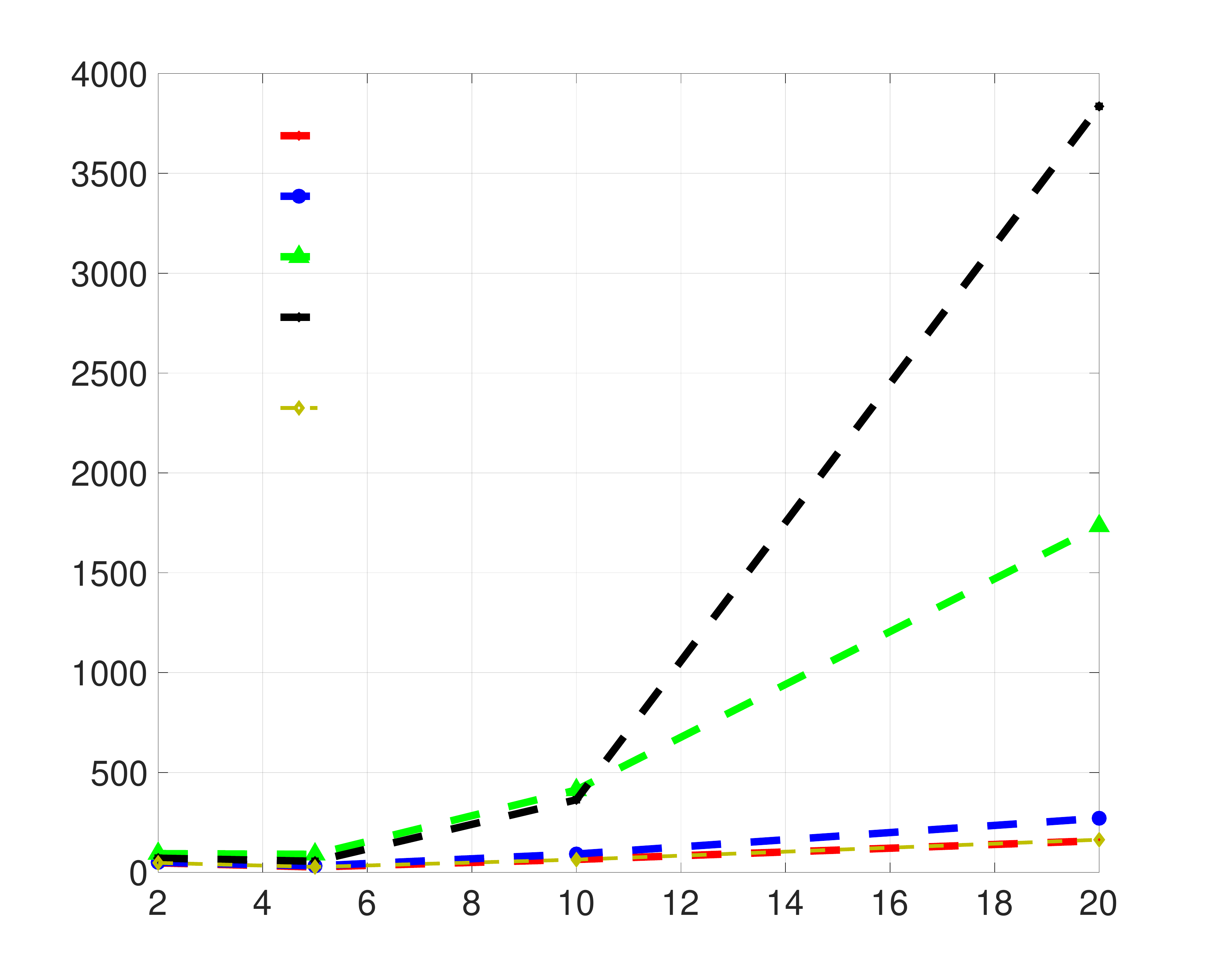}
\end{overpic}
\put(-150,142){\footnotesize{MA}}
\put(-150,131){\footnotesize{MW}}
\put(-150,121){\footnotesize{Randomized}}
\put(-150,110){\footnotesize{PF}}
\put(-150,100){\footnotesize{Optimal AoI}}
\put(-150,90){\footnotesize{(Theoretical)}}
\put(-132,0){\footnotesize{Number of UEs ($N$)}}
\put(-210,50){\rotatebox{90}{\footnotesize{Long Term Max-Age}}}
\caption{\small {Comparative Performance of the Proposed Max-Age (MA) policy with three other Scheduling Policies with varying the number of UEs for the problem in Section \ref{max_AoI}.}}
\label{AoI1}
\end{figure}

\paragraph*{Simulated Policies}
In this section, we simulate the following five scheduling policies for the downlink wireless system described in Section \ref{system_model} - 1) Max-Age Policy (\textsf{MA}) 2) Randomized Policy (\textsf{RP}) 3) Max-Weight Policy (\textsf{MW}) 4) Proportional Fair (\textsf{PF}) 5) Max-Age Policy with Throughput Constraints (\textsf{MATP}). The policy \textsf{MA} is described in Section \ref{max_AoI}. The second policy \textsf{RP} chooses a UE randomly in each time slot. The third policy \textsf{MW} was proposed in \cite{AoI_ToN} for approximately minimizing the long-term average-AoI metric. In every time slot $t$, the policy \textsf{MW} chooses the UE $i$ which maximizes the metric $p_ih_i^2(t)$ amongst all UEs. The fourth policy is the well-known Proportional Fair policy \cite{stolyar2005asymptotic}, \cite{PF} which at every slot $t$ selects the UE maximizing the metric $p_i/R_i(t)$. Here $R_i(t)$ is the exponentially-smoothed average rate, which is updated at every time slot as: $R_i(t + 1) = R_i(t) + \epsilon y_i(t)$ where $y_i(t)$ is the \emph{instantaneous throughput} to the UE\textsubscript{$i$} at slot $t$. The fifth policy (\textsf{MATP}) is described in Section \ref{throughput} of this paper.  
\paragraph*{Simulation Set-Up}
We simulate a downlink wireless network with $N$ nodes, each with a binary erasure channel. The probability of successful transmission for the $i$\textsuperscript{th} channel $p_i$ is sampled i.i.d. from a uniform distribution in $[0,1]$. Each simulation is run for $10^5$ slots, and an average of $100$ simulations is taken for the plots. For the \textsf{PF} algorithm, the value of $\epsilon$ is set to $0.1$. 
\paragraph*{Discussion} In Figure \ref{AoI1}, we have compared the performance of five different scheduling policies on the basis of long-term Max-Age in the set-up described in Section \ref{max_AoI}. The number of Type-I UEs associated with the BS has been varied from $2$ to $20$. For reference, we have also included the Theoretical Optimal value of AoI, given in Theorem \ref{opt}. As expected, we see that the performance of the Max-Age (\textsf{MA}) policy matches with the optimal value.  The Max-Weight policy performs slightly worse than the optimal MA policy. However, we find that the randomized and the \textsf{PF} policy performs very poorly in terms of the long-term peak-age metric. The bottom line is that a utility-maximizing policy (such as \textsf{PF}, which maximizes the summation of logarithmic rates of the UEs) may be far from optimality when maximizing freshness of information on the UE side.

\begin{figure}
\centering
\begin{overpic}[width=0.4\textwidth]{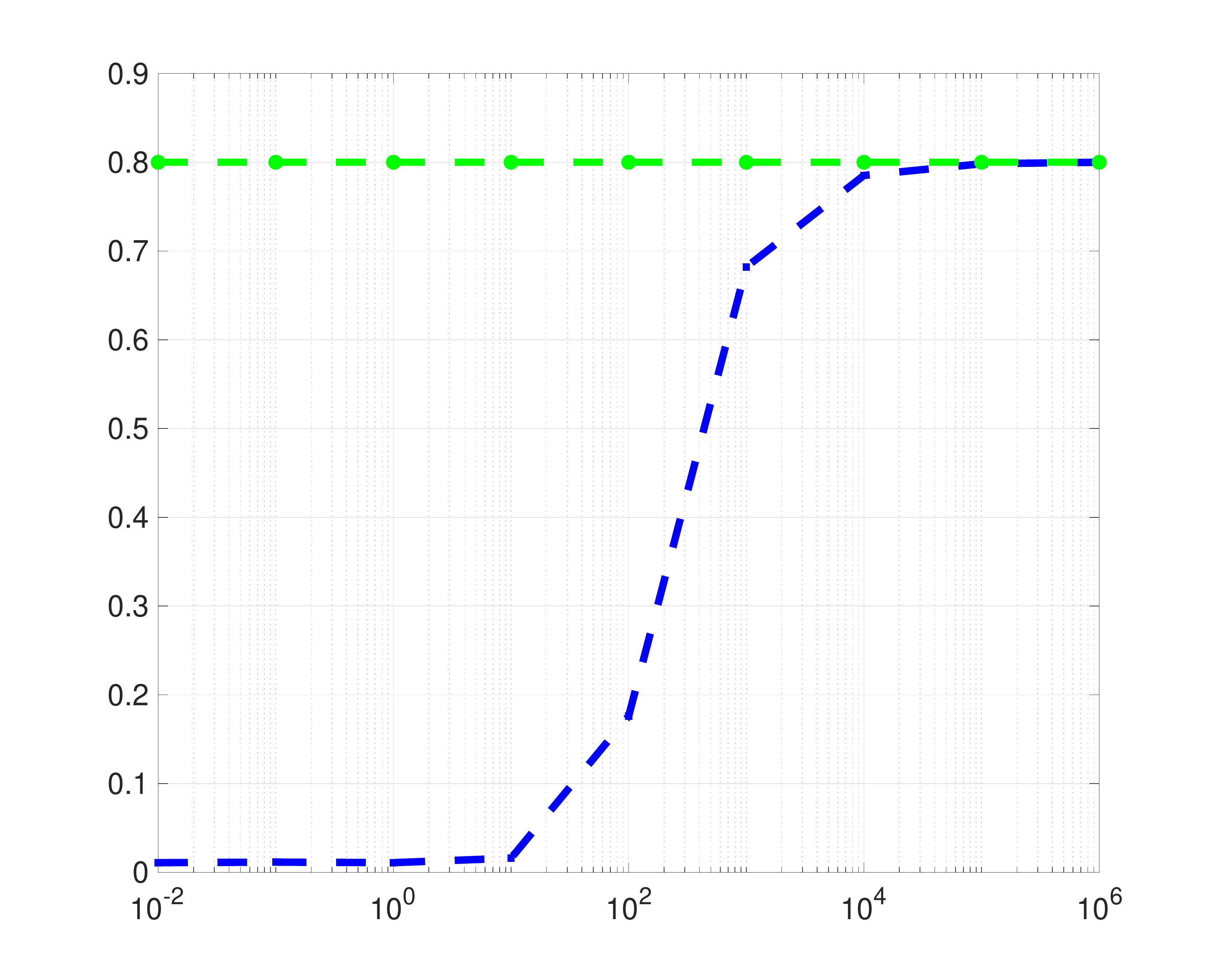}
\end{overpic}
\put(-167,125){\footnotesize{Upper-Bound of UE\textsubscript{1}}}
\put(-158, 112){\footnotesize{Throughput ($p_1$)}}
\put(-90,50){\footnotesize{Throughput of UE\textsubscript{1}}}
\put(-100,0){\footnotesize{$\beta$}}
\put(-205,44){\rotatebox{90}{\footnotesize{Throughput of UE\textsubscript{1}}}}
\caption{\small {Variation of Throughput of UE\textsubscript{1} with the parameter $\beta$.}}
\label{AoI_TPUT}
\end{figure}
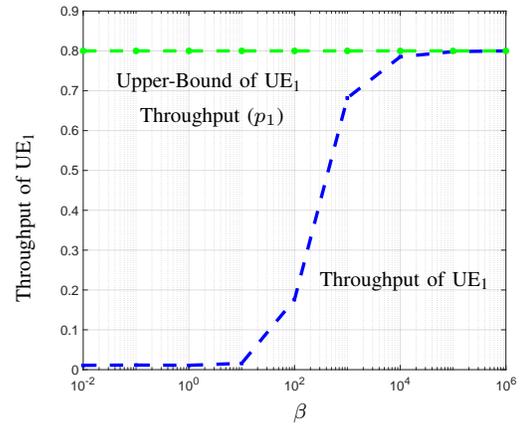

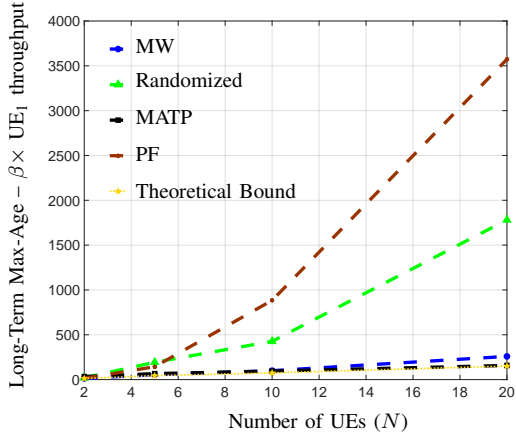
\begin{figure}
\centering
\begin{overpic}[width=0.4\textwidth]{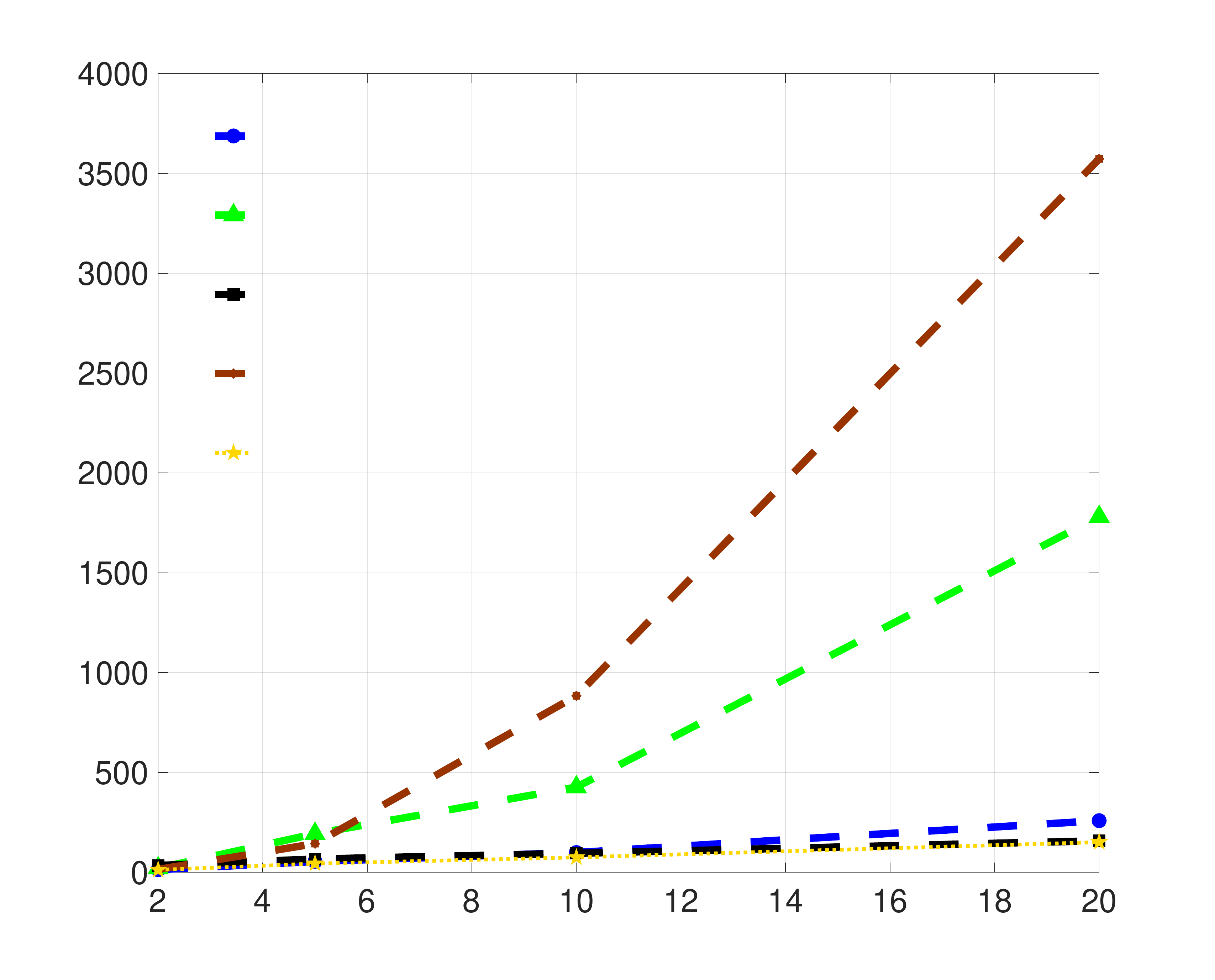}
\end{overpic}
\put(-160,142){\footnotesize{MW}}
\put(-160,129){\footnotesize{Randomized}}
\put(-160,115){\footnotesize{MATP}}
\put(-160,101){\footnotesize{PF}}
\put(-160,87){\footnotesize{Theoretical Bound}}
\put(-125,0){\footnotesize{Number of UEs ($N$)}}
\put(-208,15){\rotatebox{90}{\footnotesize{Long-Term Max-Age -- $\beta \times$ UE\textsubscript{1} throughput}}}
\caption{\small {Comparative Performance of the Proposed MATP Policy with other well-known scheduling policies for the Problem in Section \ref{throughput}.}}
\label{AoI2}
\end{figure}

Figures \ref{AoI_TPUT} and \ref{AoI2} pertain to the problem discussed in Section \ref{throughput}, where the objective is to minimize the long-term peak-AoI, while providing a certain throughput guarantee to  UE\textsubscript{1}, which serves eMBB type of traffic. Consistent with the observation that the parameter $\beta$ amplifies the cost for throughput-loss to UE\textsubscript{1} (viz. Eqn. \ref{tput}), Figure \ref{AoI_TPUT} shows that, under the action of the MATP policy,  UE\textsubscript{1} receives more throughput as the parameter $\beta$ is increased. For very large value of the $\beta$ ($\beta \approx 10^6$), the throughput to UE\textsubscript{1} saturates to $p_1$ ($p_1=0.8$ in the Figure \ref{AoI_TPUT}), which is the maximum-throughput obtainable for UE\textsubscript{1} if  UE\textsubscript{1} is scheduled exclusively. Figure \ref{AoI2} compares the performance of various scheduling policies in terms of the metric given in Eqn. \eqref{tput}. Similar to Figure \ref{AoI1}, we see that the \textsf{MW} and the Randomized Policies perform poorly in this case. However, the proposed approximately optimal policy \textsf{MATP} performs close to the theoretical bound and the performance of the \textsf{MW} policy is also not very far from that of the \textsf{MATP} policy.  


%% file: conclusion.tex
\section{Conclusion}\label{conclusion}
In this paper, we have derived an optimal downlink scheduling policy for minimizing the long-term peak-Age-of-Information for UEs with URLLC type of traffic. We have also proposed a heuristic scheduling policy in the case when one of the UEs is throughput-constrained. Extensive numerical simulations have been carried out comparing the efficacy of different scheduling policies. Deriving an optimal scheduling policy for minimizing peak-AoI with throughput constraint is an interesting research direction which will be pursued in the future.  

%% file: Acknowledgement.tex
\section{Acknowledgement}
The second author would like to thank Prof. Eytan Modiano and Igor Kadota from MIT, for the useful discussions that led to this paper.

%% file: appendix.tex
\section{Appendix}
\subsection{Proof of Proposition \ref{exp_bd}} \label{exp_bd_proof}
Using Union bound, we have 
\begin{eqnarray}\label{ub}
\mathbb{P}\big(\max_i h_i(t) \geq k\big)=\mathbb{P}(\bigcup_{i}h_i(t)\geq k)\leq 	\sum_{i}\mathbb{P}(h_i(t) \geq k).
\end{eqnarray}
Next, for any UE $i$, the event $h_i(t) \geq k$ occurs iff at time $t$, it has been at least $k$ slots since UE\textsubscript{i} received a packet successfully. Let $p_{\max}\stackrel{\textrm{(def)}}{=}\max_i p_i$ and $p_{\min}=\min_i p_i$. Since the \textsf{MA} policy successfully serves other UEs exactly once between two consecutive successful service of UE\textsubscript{i}, it follows that, during the last $k$ slots prior to time $t$, at most $N-1$ UEs have successfully received a packet. Thus, 
\begin{eqnarray*}
	&&\mathbb{P}(h_i(t) \geq k) \leq \sum_{j=0}^{N-1} \binom{k}{j}p_{\max}^{j}(1-p_{\min})^{k-j} \\
	&&\leq  \binom{k}{N-1}(1-p_{\min})^k\\
	&&\times \frac{1-p_{\min}}{(p_{\max}+p_{\min})-1}\bigg(\big(\frac{p_{\max}}{1-p_{\min}}\big)^{N}-1\bigg)\\	
&&\leq c'(N, \bm{p}) k^N (1-p_{\min})^k,
\end{eqnarray*} 
where we have used the bound $\binom{k}{N-1}\leq \frac{k^N}{(N-1)!}$ and defined $c'(N, \bm{p})\equiv \frac{1-p_{\min}}{(N-1)!((p_{\max}+p_{\min})-1)}\bigg(\big(\frac{p_{\max}}{1-p_{\min}}\big)^{N}-1\bigg) $\footnote{In the case $p_{\max}/(1-p_{\min})=1$, we take $c'\equiv N/(N-1)!$.}. The final result now follows from Eqn. \eqref{ub}.


\subsection{Proof of Theorem \ref{stability}} \label{stability_proof}
It is clear that $\{\bm{h}(t)\}_{t \geq 1}$	forms a countable-state Markov Chain under the action of the \textsf{MA} policy. To show the positive recurrence of the chain $\{\bm{h}(t)\}_{t \geq 1}$, we analyze the stochastic dynamics of the random variable $h_{\textrm{avg}}(t)\stackrel{\textrm{(def)}}{=}N^{-1}\sum_{i=1}^{N}h_i(t)$, and choose it as our Lyapunov function for the subsequent drift analysis. \\
Let $i^*(t)= \arg \max_{i} h_i(t)$, where we break ties arbitrarily. Then, the \textsf{MA} policy transmits a fresh packet to the $i^*(t)$\textsuperscript{th} user at time $t$. Over the binary erasure channel that we consider, this packet transmission is successful with probability $p_{i^*(t)}$ and is unsuccessful w.p. $1-p_{i^*(t)}$. In case the packet transmission is unsuccessful, the age of all users increase by $1$. Thus, 
\begin{eqnarray}\label{eq1}
	h_{\textrm{avg}}(t+1)=h_{\textrm{avg}}(t)+1~~\textrm{w.p.}~~ 1-p_{i^*(t)}.
\end{eqnarray}

On the other hand, in the case when the packet transmission \newpage is successful, the age of the $i^*(t)$\textsuperscript{th} user drops to $1$, and the age of all other users increases by $1$. Hence, we can write 
\begin{eqnarray*}\label{eq2}
	Nh_{\textrm{avg}}(t+1)=Nh_{\textrm{avg}}(t)+(1-h_{i^*(t)}) + N-1,~\textrm{w.p. } ~p_{i^*(t)}.
\end{eqnarray*} 
Finally, note that, under the action of the \textsf{MA} policy, we have $h_{i^*(t)}=\max_{i}h_i(t) \geq h_{\textrm{avg}}(t)$. Hence, from the above equation, we conclude that
\begin{eqnarray} \label{eq2}
	h_{\textrm{avg}}(t+1)\leq (1-\frac{1}{N})h_{\textrm{avg}}(t) + 1, ~~\textrm{w.p.} ~~ p_{i^*(t)}.
\end{eqnarray}
Let $\mathcal{F}_t$ be the sigma-field generated by the random variables $\{\bm{h}(k), 1\leq k \leq t\}$, \emph{i.e.,} $\mathcal{F}_t= \sigma(\bm{h}(1), \bm{h}(2), \ldots, \bm{h}(t)), t\geq 1.$ Using Equations \eqref{eq1} and \eqref{eq2}, we upper-bound the one-slot conditional drift $\Delta(h_{\textrm{avg}}(t))$ as follows:
\begin{eqnarray}\label{drift}
	&&\Delta(h_{\textrm{avg}}(t)) 
	\equiv \mathbb{E}(h_{\textrm{avg}}(t+1)-h_{\textrm{avg}}(t)|\mathcal{F}_t)\nonumber \\
	&\leq & (1-p_{i^*(t)})(h_{\textrm{avg}}(t)+1)+p_{i^*(t)}\bigg((1-\frac{1}{N})h_{\textrm{avg}}(t) + 1\bigg)\nonumber \\
	&&-h_{\textrm{avg}}(t)\nonumber \\
	&=& 1-\frac{p_{i^*(t)}}{N}h_{\textrm{avg}}(t)
	\leq  1-\frac{p_{\min}}{N}h_{\textrm{avg}}(t)\nonumber \\
	&=& 1-\frac{p_{\min}}{N^2}\sum_{i=1}^{N}h_i(t),
\end{eqnarray}
where $p_{\min}=\min_i p_i >0$. The drift upper-bound \eqref{drift} shows that if $h_k(t) \geq \frac{2N^2}{p_{\min}}$ for any $1\leq k \leq N$, we have $\Delta(h_{\textrm{avg}}(t)) \leq -1 <0$. Thus, the one-slot conditional drift of the chosen Lyapunov function is strictly negative whenever the state $\bm{h}(t)$ lies outside the bounded $N$-dimensional box $[0, \frac{2N^2}{p_{\min}}]^N$. Finally, using the Foster-Lyapunov Theorem for stability of Markov Chains (Proposition 6.13 (b) of \cite{hajek2015random}), we conclude that the Markov Chain $\{\bm{h}(t)\}_{t \geq 1}$ is Positive Recurrent.  
